\documentclass{vgtc}

\graphicspath{{figures/}{pictures/}{images/}{./}}

\usepackage{times}

\usepackage{amsmath}

\usepackage{mathptmx}
\usepackage{amssymb}
\usepackage{array}
\usepackage[table]{xcolor}
\usepackage{enumitem}

\onlineid{0000}

\vgtccategory{Research}

\vgtcinsertpkg

\nocopyrightspace

\newcommand\blfootnote[1]{%
  \begingroup
  \renewcommand\thefootnote{}\footnote{#1}%
  \addtocounter{footnote}{-1}%
  \endgroup
}

\title{VocabulARy 2.0: A Generative AR System Enabling Personalised 3D Mnemonics for Vocabulary Learning in Context} 

\author{Karolina Trajkovska\thanks{e-mail: karolina.trajkovska@famnit.upr.si}\\ %
        \parbox{1.4in}{\scriptsize \centering University of Primorska} %
\and Matja{\v{z}} Kljun\thanks{e-mail: matjaz.kljun@upr.si}\\ %
     \parbox{1.4in}{\scriptsize \centering University of Primorska \\ Stellenbosch University}        
\and Maheshya Weerasinghe\thanks{e-mail: maheshya.weerasinghe@famnit.upr.si}\\ %
     \parbox{1.4in}{\scriptsize \centering University of Primorska} %
\and Klen {\v{C}}opi{\v{c}} Pucihar\thanks{e-mail: klen.copic@famnit.upr.si}\\ %
    \parbox{1.4in}{\scriptsize \centering University of Primorska \\ Stellenbosch University}}

\teaser{
  \centering
  \includegraphics[width=\linewidth]{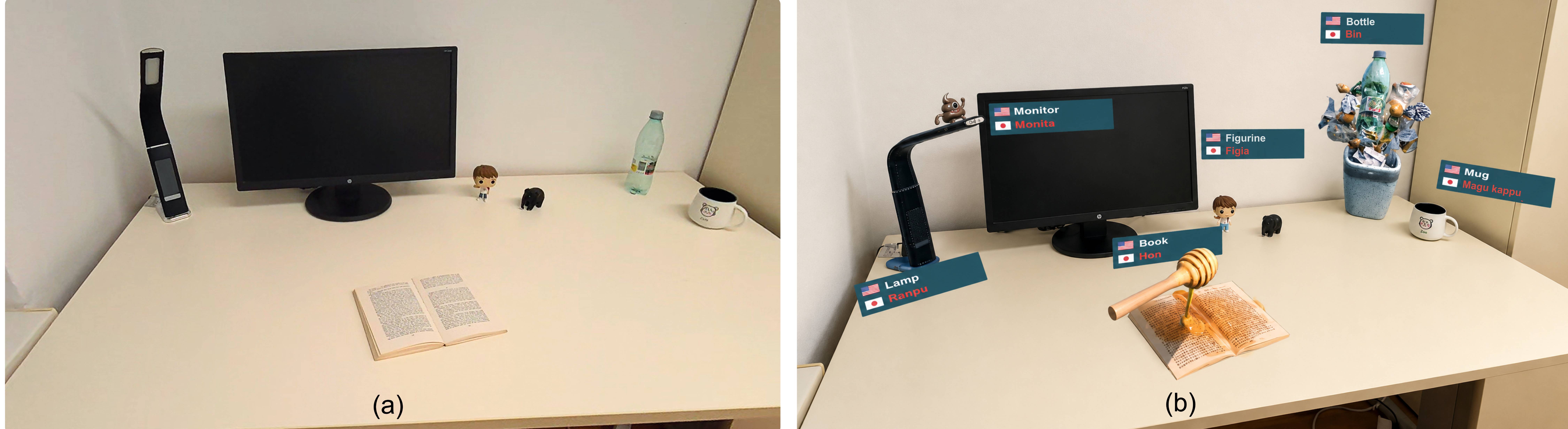}
  \caption{Real-world workspace before (a) and after (b) personalised mnemonic augmentation. Note: Image (b) was minimally edited for presentation purposes, primarily to improve text readability.}
  \label{fig:teaser}
}

\abstract{

    Mnemonics support vocabulary learning by linking unfamiliar words to familiar concepts through memorable verbal and visual associations. We present a generative AR system that enables learners to create personalised 3D mnemonics by transforming familiar objects in their surroundings into situated mnemonic anchors. The system scans the environment, detects visible objects, and generates bilingual flashcards. After selecting a word, the learner describes a personal association. The system combines this description with an
    image of the selected object to generate a textured 3D model using generative AI. Learners can then place the model relative to the real object, transforming it into a situated mnemonic anchor. 
} 


\begin{document}

\firstsection{Introduction}

\maketitle

\blfootnote{%
  \noindent\scriptsize\itshape
  \rule{0pt}{3ex} Authors version. Contact author: karolina.trajkovska@famnit.upr.si\par
}

Mnemonics can support vocabulary learning by linking unfamiliar words to memorable verbal or visual cues. However, constructing such cues can introduce additional demands: alongside learning a new word, learners must understand and apply a mnemonic technique. This effort may divert attention from the primary task of vocabulary learning, motivating systems that automate or scaffold mnemonic creation. 

Early systems
generated candidate mnemonic keywords~\cite{savva2014transphoner}, while later approaches, including SmartPhone~\cite{lee2023smartphone} and recent LLM-based systems~\cite{el2024keymagine}, generated both verbal and visual cues. Other work has explored personalised and learner--AI co-created mnemonics~\cite{attygalle2025text,shao2026wordcraft}. In augmented reality~(AR), VocabulARy~\cite{weerasinghe2022vocabulary} placed bilingual labels, audio, keywords, and 3D visuals beside real-world objects, demonstrating how mnemonic support can be integrated into the learner's immediate context. However, in the case of VocabulARy, these visual cues and 3D assets must be prepared in advance, limiting their use in unprepared environments and their ability to reflect learners' own ideas.
To address this gap, we present \textit{VocabulARy 2.0}, a generative AR system that enables learners to create personalised 3D mnemonics 
by transforming familiar objects in their surroundings into situated mnemonic anchors (\autoref{fig:teaser}). 
The system scans the environment, detects physical objects, produces translated vocabulary flashcards, and allows learners to describe and generate a 3D mnemonic for a selected word.

\section{System Design and Implementation}

\begin{figure}[h!]
    \centering
    \includegraphics[width=\linewidth]{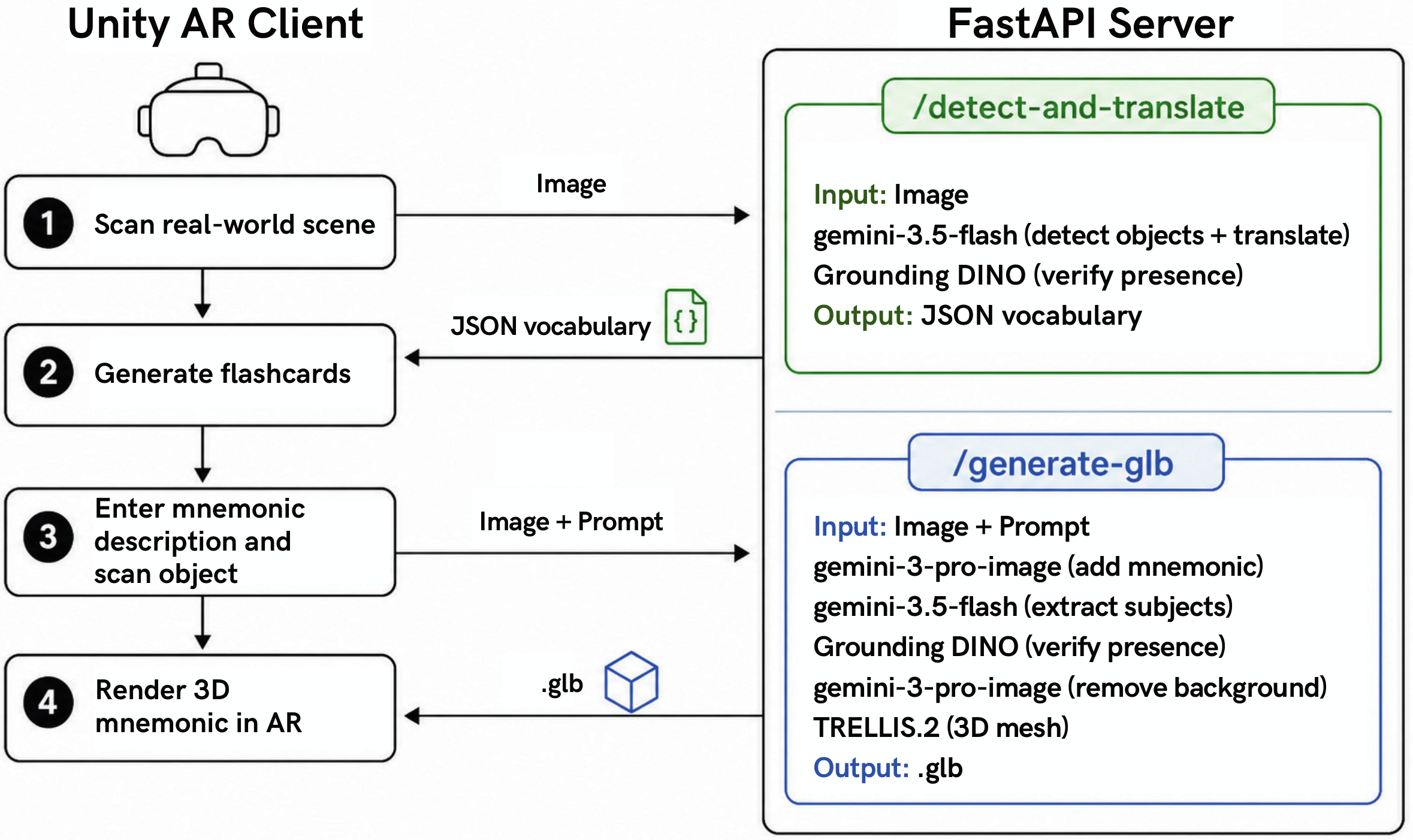}
    \caption{System architecture of \textit{VocabulARy 2.0}.}
    \label{fig:system_architecture}
\end{figure}

The system adopts a decoupled client--server architecture comprising a \textit{Unity AR Client} and a \textit{FastAPI Server} (\autoref{fig:system_architecture}). The Unity client provides the AR interface, captures images of the physical environment, manages user interaction, and renders generated 3D content. Computationally intensive computer vision and generative AI tasks are offloaded to the FastAPI server, which coordinates cloud-based Gemini models
with locally deployed Grounding DINO and TRELLIS.2 models~\cite{xiang2025trellis2}.

The workflow consists of four stages: (1) scanning a real-world scene, (2) generating bilingual flashcards, (3) entering a description and scanning the selected object, and (4) manipulating and rendering the generated 3D mnemonic in AR. Communication between the client and the server is handled through two REST API endpoints: \texttt{/detect-and-translate}, which returns detected vocabulary as JSON, and \texttt{/generate-glb}, which returns a textured 3D model as a \texttt{.glb} file.

\subsection{Unity AR Client}

The client application is developed in Unity 3D using the Meta XR SDK and runs on a stand-alone Meta Quest 3 headset.

\textbf{Stage 1: Scanning a real-world scene.}
The client uses the headset’s passthrough camera API to capture the current view of the physical environment from the GPU render texture. The captured frame is encoded as a JPEG byte array and sent to the server’s \texttt{/detect-and-translate} endpoint. The endpoint returns a JSON vocabulary response containing each detected object’s English name and its translation in the target language.

\textbf{Stage 2: Generating flashcards.}
The client uses the returned JSON vocabulary to dynamically generate one bilingual flashcard for each detected object. The flashcards are arranged in a responsive 2D grid positioned in 3D space above the workspace to minimise occlusion of physical objects.

\begin{figure}[h!]
    \centering
    \includegraphics[width=\linewidth]{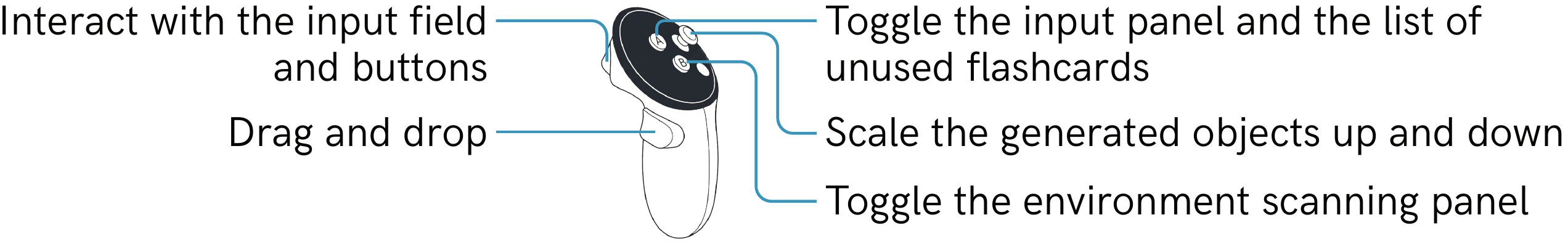}
    \caption{Controller mappings for interacting with the system.}
    \label{fig:interaction}
\end{figure}

\textbf{Stage 3: Entering a mnemonic.}
After selecting a flashcard, the learner enters a personal mnemonic description and captures an image of the corresponding real-world object. The image and textual description are then sent to the \texttt{/generate-glb} endpoint.
Interface elements, including flashcards and input panels, use box colliders and can be grabbed and repositioned using the headset controllers, as illustrated in \autoref{fig:interaction}.

\textbf{Stage 4: 3D model manipulation and rendering.}
The generated \texttt{.glb} file (the result of a prior request to the \texttt{/generate-glb} endpoint) is downloaded and instantiated asynchronously in the active Unity scene using the glTFast library, without requiring the application to be rebuilt. The learner can resize, reposition, and align the model with the corresponding physical object, thereby transforming it into a situated mnemonic anchor.

\subsection{FastAPI Server}

The backend is implemented using FastAPI and runs on a local workstation equipped with two NVIDIA RTX PRO 6000 Blackwell GPUs with 96~GB of VRAM each and an AMD Ryzen Threadripper PRO 7975WX 32-core processor.
The server provides two processing pipelines.

\textbf{Vision and translation pipeline (\texttt{/detect-and-translate}).}
This endpoint receives an image of the learner's environment. The \texttt{gemini-3.5-flash} model identifies visible objects and translates their English names into the selected target language. A locally deployed Grounding DINO model then verifies that the identified objects are present in the image. The endpoint returns a JSON vocabulary mapping each verified English term to its target-language translation, which the Unity client uses to generate the flashcards.

\textbf{3D generative pipeline (\texttt{/generate-glb}).}
This endpoint receives the learner's mnemonic description together with an image of the selected object. The pipeline first uses \texttt{gemini-3-pro-image} to incorporate the described mnemonic into the image. The relevant subjects are then extracted using \texttt{gemini-3.5-flash}, and a locally deployed Grounding DINO verifies their presence. A second \texttt{gemini-3-pro-image} step removes the background and produces a clean image of the processed subjects. This image is passed to a locally deployed TRELLIS.2 instance, which generates a textured 3D mesh. The resulting asset is packaged as a \texttt{.glb} file and returned to the Unity client for manipulation and rendering in AR.

The architecture uses a hybrid local and cloud-based pipeline. Grounding DINO and TRELLIS.2 run locally for object verification and 3D mesh generation, whereas Gemini APIs are used for image understanding and synthesis because the locally evaluated alternatives (Qwen/Qwen-Image-Edit-2511, timbrooks/instruct-pix2pix and black-forest-labs/FLUX.1-dev) did not provide comparable quality and consistency for the required transformations.

\section{Limitations and Conclusion}

\textit{VocabulARy 2.0} demonstrates an end-to-end workflow for generating personalised 3D vocabulary mnemonics in unprepared environments. However, the current prototype requires $\bar{x} = 1\text{min } 55\text{s}, \quad SD = 32\text{s}$ to generate each mnemonic.
Given the generation time, the demo will include pre-generated mnemonics on a headset and tablets, enabling attendees to interact with them immediately in AR.
Another limitation of the current implementation
is that the generated models do not have optimised mesh which
causes rendering issues (flicker and tracking issues) on the Meta Quest 3 when three or more models are viewed simultaneously. Future work should therefore focus on reducing generation latency and optimising meshes.
Mnemonic descriptions are currently entered using a virtual keyboard; future work should explore speech input to reduce interaction effort in AR. Further studies are also needed to evaluate usability, cognitive load, and vocabulary learning outcomes.

\section{GenAI Usage Disclosure}

We used ChatGPT and Gemini for correcting grammar, enhancing expressions, and programming assistance.

\bibliographystyle{abbrv-doi}

\bibliography{template}
\end{document}